\documentclass[twocolumn,english,aps,prb,twocolum,superscriptaddress,bibnotes,amsmath,amssymb,floatfix]{revtex4-1}
\usepackage[colorlinks=true,citecolor=blue,linkcolor=magenta]{hyperref}

\usepackage[markup=nocolor, authormarkupposition=left]{changes} 
\usepackage{soul}
\usepackage[utf8]{inputenc}
\usepackage[english]{babel}
\usepackage{amsmath,amsfonts,amssymb}
\usepackage[T1]{fontenc}
\usepackage{url}
\usepackage{changes}

\usepackage{amsmath}
\usepackage{amsfonts}
\usepackage{amssymb}
\usepackage{siunitx}
\usepackage{epstopdf}
\usepackage{graphicx}
\graphicspath{{./Figures/}}

\begin{document}
\title{High-yield wafer-scale fabrication of ultralow-loss, dispersion-engineered\\silicon nitride photonic circuits}

\author{Junqiu Liu}
\affiliation{Institute of Physics, Swiss Federal Institute of Technology Lausanne (EPFL), CH-1015 Lausanne, Switzerland}

\author{Guanhao Huang}
\affiliation{Institute of Physics, Swiss Federal Institute of Technology Lausanne (EPFL), CH-1015 Lausanne, Switzerland}

\author{Rui Ning Wang}
\affiliation{Institute of Physics, Swiss Federal Institute of Technology Lausanne (EPFL), CH-1015 Lausanne, Switzerland}

\author{Jijun He}
\affiliation{Institute of Physics, Swiss Federal Institute of Technology Lausanne (EPFL), CH-1015 Lausanne, Switzerland}

\author{Arslan S. Raja}
\affiliation{Institute of Physics, Swiss Federal Institute of Technology Lausanne (EPFL), CH-1015 Lausanne, Switzerland}

\author{Tianyi Liu}
\affiliation{Institute of Physics, Swiss Federal Institute of Technology Lausanne (EPFL), CH-1015 Lausanne, Switzerland}

\author{Nils J. Engelsen}
\affiliation{Institute of Physics, Swiss Federal Institute of Technology Lausanne (EPFL), CH-1015 Lausanne, Switzerland}

\author{Tobias J. Kippenberg}
\email[]{tobias.kippenberg@epfl.ch}
\affiliation{Institute of Physics, Swiss Federal Institute of Technology Lausanne (EPFL), CH-1015 Lausanne, Switzerland}

\maketitle
\noindent\textbf{
Low-loss photonic integrated circuits (PIC) and microresonators have enabled novel applications ranging from narrow-linewidth lasers \cite{HuangD:19, Xiang:20}, microwave photonics \cite{Marpaung:19, Eggleton:19}, to chip-scale optical frequency combs \cite{Kippenberg:18, Gaeta:19} and quantum frequency conversion \cite{Li:16, Lu:19}. 
To translate these results into a widespread technology \cite{Agrell:16}, attaining ultralow optical losses with established foundry manufacturing is critical. 
Recent advances in fabrication of integrated Si$_3$N$_4$ photonics \cite{Xuan:16, Ji:17, Liu:18a, Ye:19, ElDirani:19} have shown that ultralow-loss, dispersion-engineered microresonators can be attained at \textit{die-level} throughput. 
For emerging nonlinear applications such as integrated travelling-wave parametric amplifiers \cite{Foster:06, Kuyken:11, Ooi:17, YangM:18} and mode-locked lasers \cite{Xin:19}, PICs of length scales of up to a meter are required, placing stringent demands on yield and performance that have not been met with current fabrication techniques.
Here we overcome these challenges and demonstrate a fabrication technology which meets all these requirements on \textit{wafer-level} yield, performance and length scale.  
Photonic microresonators with a \emph{mean} $Q$ factor exceeding $30\times10^6$, corresponding to a linear propagation loss of 1.0 dB/m, are obtained over full 4-inch wafers, as determined from a statistical analysis of tens of thousands of optical resonances and cavity ringdown with 19 ns photon storage time.
The process operates over large areas with high yield, enabling 1-meter-long spiral waveguides with 2.4 dB/m loss in dies of only $5\times5$ mm$^2$ size.
Using a modulation response measurement self-calibrated via the Kerr nonlinearity, we reveal that, strikingly, the intrinsic absorption-limited $Q$ factor of our Si$_3$N$_4$ microresonators exceeds $10^9$. 
This absorption loss is sufficiently low such that the Kerr nonlinearity dominates the microresonator's modulation response even in the audio frequency band. 
Transferring the present Si$_3$N$_4$ photonics technology to standard commercial foundries, and merging it with silicon photonics using heterogeneous integration technology \cite{Liang:10a, Park:20, Blumenthal:20}, will significantly expand the scope of today's integrated photonics and seed new applications.
}

Silicon photonics \cite{Thomson:16} has evolved into a mature technology enabling the generation, modulation and detection of optical signals on-chip, via heterogeneous or hybrid integration of different material platforms \cite{Liang:10a, Park:20, Blumenthal:20}. 
Within the past two decades, it has been transferred from academic research to large-volume commercial deployment in datacenter interconnects.
A second revolution is currently under way in which, the optical nonlinearities of PIC - accessed with continuous-wave lasers at sub-milliwatt power - become relevant for applications, i.e. \emph{integrated nonlinear photonics}.
The Kerr, $\chi^{(2)}$ or Brillouin nonlinearities enable novel schemes for nonlinear optical signal generation and processing \cite{Marpaung:19, Gaeta:19, WangC:18}. 
Major effort has been made in the past decade in developing various integrated nonlinear photonic platforms ranging from Si$_3$N$_4$ \cite{Xuan:16, Ji:17, Liu:18a, Ye:19, ElDirani:19}, diamond \cite{Hausmann:14}, Ta$_2$O$_5$ \cite{Jung:19}, SiC \cite{Lukin:20} to highly nonlinear AlGaAs \cite{Pu:16, Chang:20} and GaP \cite{Wilson:20} on insulator, as well as electro-optic platforms such as LiNbO$_3$ \cite{WangC:18, Zhang:19, He:19, Fang:19} and AlN \cite{Jung:14, GuoX:16, LiuX:18, LiuX:18a}. 
Significant progress has been achieved on harnessing the Kerr nonlinearity which enables the generation of dissipative Kerr soliton microcombs in integrated optical microresonators \cite{Kippenberg:18, Gaeta:19}. 
Microcombs constitute chip-scale frequency combs with broad bandwidths and repetition rates in the microwave domain, amenable to heterogeneous or hybrid integration with III-V/Si lasers \cite{Stern:18, Raja:19}, and have been used in system-level demonstrations including coherent telecommunications \cite{Marin-Palomo:17}, integrated frequency synthesizers \cite{Spencer:18},  astronomical spectrometer calibration \cite{Obrzud:19, Suh:19}, ultrafast ranging \cite{Trocha:18, Suh:18}, low-noise microwave generation \cite{Liang:15, Liu:20} and massively parallel coherent LiDAR \cite{Riemensberger:20}. 

\begin{figure*}[t!]
\centering
\includegraphics[clip,scale=1]{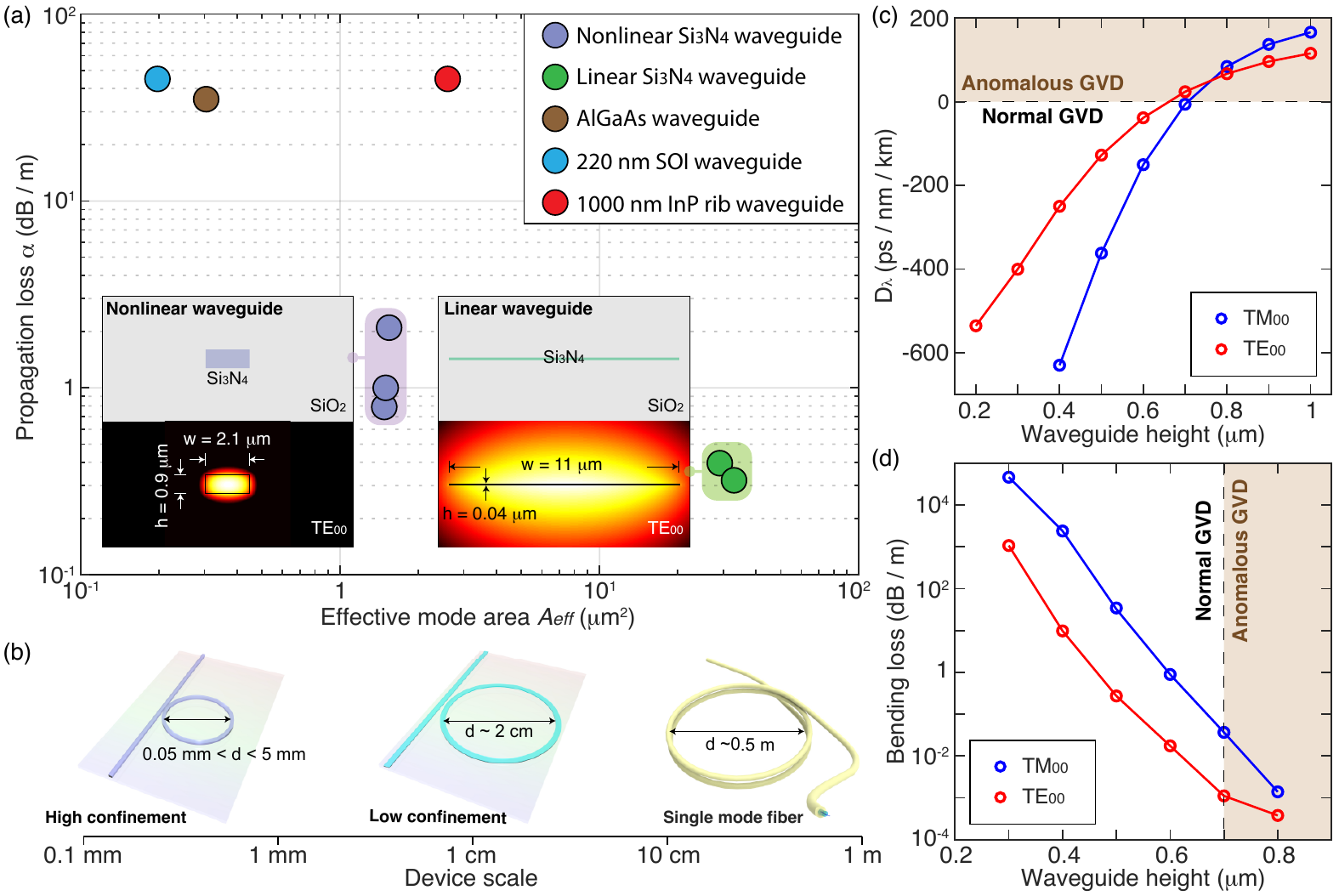}
\caption{
\textbf{Comparison of ultralow-loss linear and nonlinear Si$_3$N$_4$ platforms with state-of-the-art silicon, InP and AlGaAs platforms.}
(a)~Comparison of optical losses and effective mode areas (in the telecommunication band of 1550 nm) in state-of-the-art, lowest-loss waveguides, including nonlinear (ref. \cite{Xuan:16, Ji:17} and this work) and linear (ref. \cite{Spencer:14, Gundavarapu:19}) Si$_3$N$_4$ waveguides, 220 nm silicon-on-insulator (SOI) waveguides \cite{Selvaraja:14}, 1000 nm InP rib waveguides \cite{Ciminelli:13}, nonlinear AlGaAs waveguides \cite{Chang:20}.
The insets show the waveguide geometry and optical mode profile of the Si$_3$N$_4$ waveguides.
(b)~Comparison in device size for linear and nonlinear Si$_3$N$_4$ waveguides and single-mode fibers. 
(c)~Simulated GVD parameter $D_\lambda$ as a function of the waveguide height $h$, with a fixed waveguide width of $w=2.1$ $\mu$m.
(d)~Simulated bending loss as a function of the waveguide height, with a fixed bending radius of $d/2=25$ $\mu$m and waveguide width of $w=2.1$ $\mu$m.
Anomalous GVD region is brown-shaded, which is accessed with $h>700$ nm.
}
\label{Fig:Fig0}
\end{figure*}

For nonlinear integrated photonics, Si$_3$N$_4$ \cite{Levy:10, Moss:13, Blumenthal:18} has emerged as a leading material due to its ultralow linear and nonlinear optical losses, strong Kerr nonlinearity, high refractive index, and high power handling capability \cite{Gyger:19}. 
To date, among all integrated photonics platforms \cite{Kovach:20}, optical losses near or below 1 dB/m have only been demonstrated in Si$_3$N$_4$ waveguides.   
First achieved in thin-core waveguides (e.g. waveguide height $h<100$ nm) \cite{Spencer:14, Gundavarapu:19, Bauters:11}, ultralow losses have later also been attained in thick-core (i.e. $h>700$ nm) \cite{Xuan:16, Ji:17, Liu:18a} waveguides enabling negligible bending loss, dispersion engineering and significantly higher Kerr nonlinearity, as outlined in Fig. \ref{Fig:Fig0}. 
Many system-level demonstrations of soliton microcombs \cite{Marin-Palomo:17, Obrzud:19, Trocha:18, Liu:20, Riemensberger:20} have been based on this type of Si$_3$N$_4$ PICs.
Figure \ref{Fig:Fig0}(a) highlights the lowest-loss nonlinear (ref. \cite{Xuan:16, Ji:17} and this work) and linear \cite{Spencer:14, Gundavarapu:19} Si$_3$N$_4$ waveguides in terms of their optical losses and effective area of the fundamental optical mode, in comparison with the state-of-the-art, lowest-loss silicon \cite{Selvaraja:14}, InP \cite{Ciminelli:13} and AlGaAs \cite{Chang:20} waveguides. 
The tight confinement significantly relaxes the bending loss, a key parameter for device footprint and photonic integration, as outlined in Fig. \ref{Fig:Fig0}(b). 
Though the desirable combination of tight confinement, ultralow loss and anomalous GVD has been achieved to date \cite{Xuan:16, Ji:17}, it has only been attained in individual chips, i.e. with \textit{die-level} throughput. 
Meanwhile, the fabrication of densely packed, meter-long PIC has not been achieved.  
Nor has \textit{wafer-level} fabrication yield, reliability and reproducibility, required for widespread adoption in CMOS foundries, been demonstrated. 
Yet, densely packed, meter-long nonlinear Si$_3$N$_4$ PIC could enable a new class of devices, ranging from integrated travelling-wave parametric amplifiers \cite{Foster:06, Kuyken:11, Ooi:17, YangM:18} to integrated mode-locked-lasers based on rare-earth doping \cite{Xin:19}.

\begin{figure*}[t!]
\centering
\includegraphics[clip,scale=1]{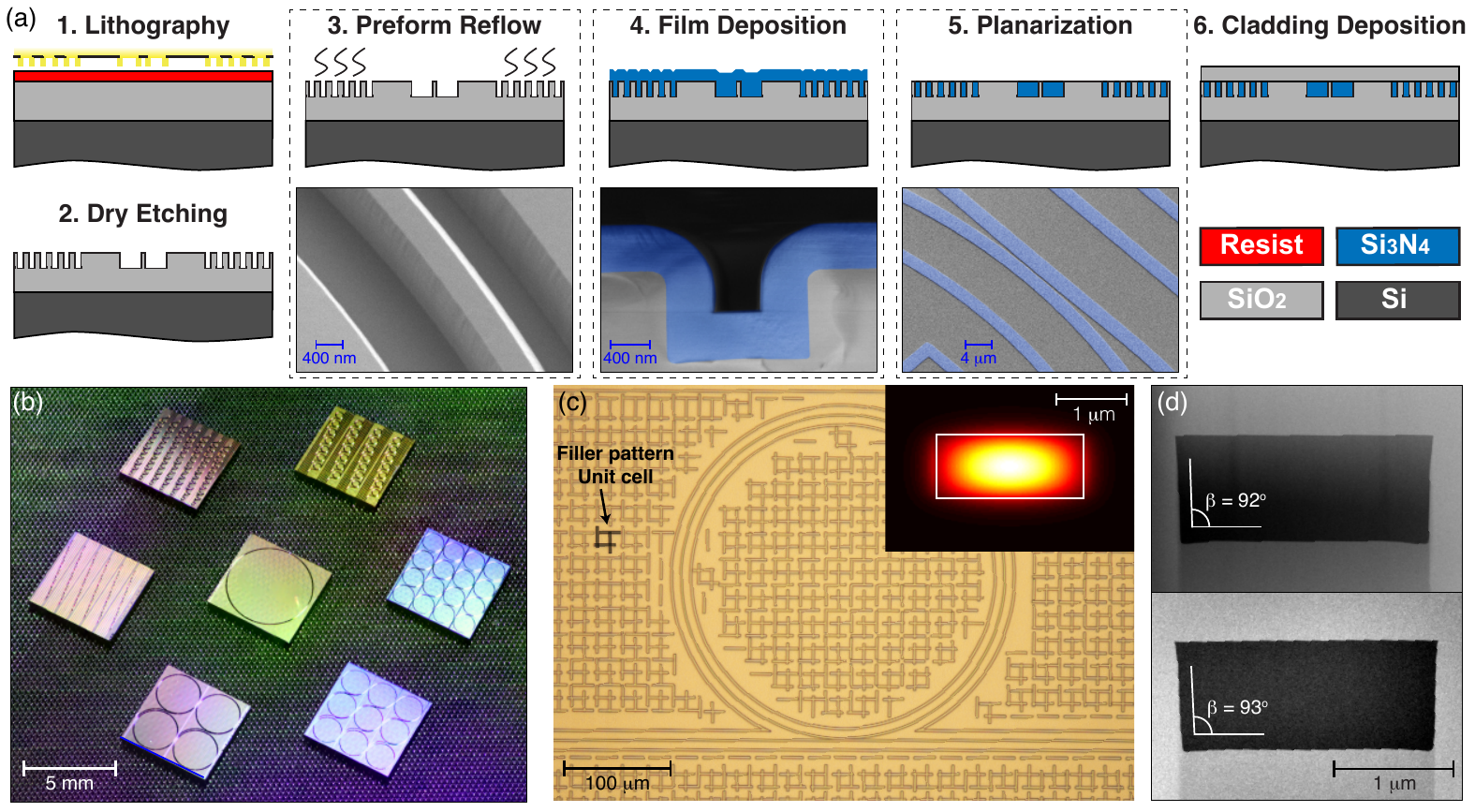}
\caption{
\textbf{The photonic Damascene process flow and highlighted features}. 
(a)~Process flow of the photonic Damascene process including DUV stepper lithography, preform etching, preform reflow, LPCVD Si$_3$N$_4$ deposition, planarization, and SiO$_2$ cladding deposition. 
The blue shaded part is Si$_3$N$_4$.
(b)~Photograph showing Si$_3$N$_4$ photonic chips with ring resonators of 10 to 1000 GHz free spectral range. 
(c)~Optical micrograph showing the bus waveguide, microring resonator, and filler patterns (used to prevent crack formation).
Inset: simulated tightly confined optical mode.
(d)~Transmission electron micrographs (TEM) of the waveguide cross-sections, before (top) and after (bottom) the preform reflow. 
The reflow preserves the waveguide dimensions accurately, while removing high-frequency spatial roughness.
}
\label{Fig:Fig1}
\end{figure*}

Here we report a high-yield wafer-scale fabrication technology to build tight-confinement, ultralow-loss, dispersion-engineered Si$_3$N$_4$ waveguides of length scales up to more than a meter.
It is based on the photonic Damascene process \cite{Pfeiffer:18b} using standard CMOS fabrication techniques such as DUV stepper lithography, dry etching and low-pressure chemical vapor deposition (LPCVD).
Figure \ref{Fig:Fig1}(a) shows  process flow and scanning electron micrographs (SEM) for selected key steps. 
The waveguides and stress-release filler patterns \cite{Pfeiffer:18b} are written directly on the SiO$_2$ substrate via DUV stepper lithography based on 248 nm KrF excimer laser. 
The use of DUV, in contrast to the commonly employed electron-beam lithography, enables dramatic increase in throughput, stability and reproducibility, essential to large-volume manufacturing.
The patterns are then dry-etched to the SiO$_2$ substrate to create waveguide preforms. 
We note that our SiO$_2$ dry etching does not introduce a trade-off between the etch verticality and surface roughness. 
Figure \ref{Fig:Fig1}(d) top shows the sidewall bottom angle $90^\circ<\beta<92^\circ$. 
To further reduce the waveguide sidewall roughness (root mean square) to sub-nanometer level, the entire substrate is annealed at 1250$^\circ$C (``preform reflow'') \cite{Pfeiffer:18}.
Importantly, this reflow process can further reduce the scattering loss, and does not lead to prominent deformation of the waveguide preform.
Figure \ref{Fig:Fig1}(d) bottom shows the measured sidewall bottom angle $\beta\approx93^\circ$. 
An LPCVD Si$_3$N$_4$ film of 1000 nm thickness is deposited on the patterned substrate, filling the preform trenches and forming the waveguides. 
A novel etchback planarization process is applied, combining photoresist coating, dry etching and chemical-mechanical planarization (CMP). 
This process enables full control of polishing depth and wafer-scale uniformity with variation below 3\%.
Afterwards, the entire substrate is thermally annealed at 1200$^\circ$C to drive out the residual hydrogen impurities in the Si$_3$N$_4$ film \cite{Liu:18a}.
Top SiO$_2$ cladding composed of TEOS and low-temperature oxide (LTO) are deposited on the wafer, followed by SiO$_2$ thermal annealing.
Finally, the wafer is separated into chips via deep dry etching followed by dicing or backside grinding, to attain chip facets with superior quality which is critical for edge coupling \cite{Raja:19, Shen:19, Voloshin:19}.

Figure \ref{Fig:Fig1}(b) shows the final Si$_3$N$_4$ chips containing multiple ring resonators of different free spectral ranges (FSR).
Figure \ref{Fig:Fig1}(c) shows the optical micrograph of the Si$_3$N$_4$ ring resonator, bus waveguide and filler patterns, as well as the tightly confined waveguide mode.
The resulted negligible bending loss allows microresonators of small radii below 23 $\mu$m (i.e. 1 THz FSR), which find wide applications in optical filters and coupled-resonator-based delay lines \cite{Poon:04, Cardenas:10}. 
The filler patterns consist of horizontal and vertical bars uniformly distributed over the entire wafer area, and can significantly relax the as-deposited LPCVD Si$_3$N$_4$ film stress for crack prevention.
These filler patterns are also required for etching and CMP uniformity.

\begin{figure*}[t!]
\centering
\includegraphics[clip,scale=1]{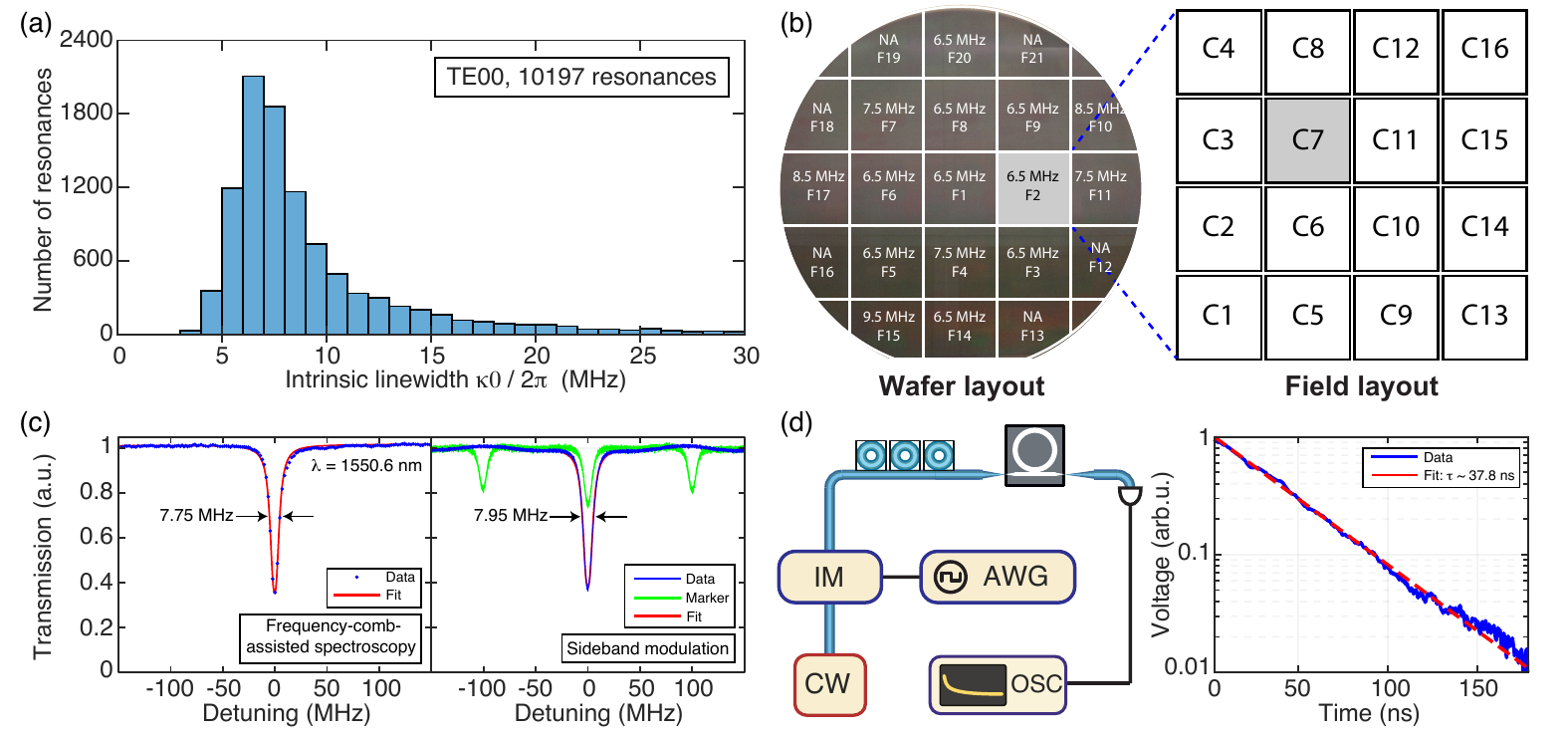}
\caption{
\textbf{Statistic study of microresonator $Q$ factors using multiple techniques}. 
(a) Histogram of 10,197 TE$_{00}$ resonances from twenty-six resonators, showing the most probable value of $\kappa_0/2\pi=6.5$ MHz and $Q_0=30\times10^6$. 
(b) DUV stepper exposure layout, and the most probable value $\kappa_0/2\pi$ of the C7 chips at different positions on the wafer.
The reticle design containing sixteen chips is uniformly exposed in discrete fields over a 4-inch wafer. 
NA: not applicable, due to visible photoresist coating defects or missing C7 chips in fields near the wafer edge.
(c) Linewidth measurement of the same resonance at 1550.6 nm using frequency-comb-assisted diode laser spectroscopy (left, $\kappa/2\pi=7.75$ MHz and $\kappa_0/2\pi=5.87$ MHz) and sideband modulation technique (right, $\kappa/2\pi=7.95$ MHz and $\kappa_0/2\pi=6.05$ MHz). 
This resonance does not present a visible mode split.
(d) Cavity ring-down measurement. 
An intensity modulator (IM) is used to switch off the pump field. 
The cavity ring-down signal is averaged 1000 times. 
The exponential fit gives an optical field decay time of $\tau=37.8$ ns, corresponding to a photon decay time of 18.9 ns and a loaded linewidth of $\kappa/2\pi=8.4$ MHz.
arb.u.: arbitrary unit.
AWG: arbitrary function generator.
OSC: oscilloscope. 
}
\label{Fig:Fig2}
\end{figure*}

\noindent \textbf{Statistical analysis of microresonator $Q$ factors}:
We fabricate Si$_3$N$_4$ microresonators of 40.6 GHz FSR, 2200 nm width and 950 nm height, and systematically study the microresonator $Q$ factors (i.e. loss).
Frequency-comb-assisted diode laser spectroscopy \cite{Delhaye:09, Liu:16} is used to characterize the resonance frequency $\omega/2\pi$ and linewidth $\kappa/2\pi$, which relate to the resonance $Q$ factor as $Q=\omega/\kappa$. 
Here we mainly study the fundamental transverse electric (TE$_{00}$) mode.
The total (loaded) linewidth $\kappa/2\pi=(\kappa_0+\kappa_\text{ex})/2\pi$, the intrinsic loss $\kappa_0/2\pi$ and the coupling strength $\kappa_\text{ex}/2\pi$ are extracted from each resonance fit. 
Figure \ref{Fig:Fig2}(a) shows the $\kappa_0/2\pi$ histogram of 10,197 TE$_{00}$ resonances measured from twenty-six microresonators. 
The most probable value is $\kappa_0/2\pi=6.5$ MHz, corresponding to an intrinsic $Q$ factor of $Q_0=30\times10^6$. 
In comparison, $\kappa_0/2\pi=9.5$ MHz is found for the fundamental transverse magnetic (TM$_{00}$) mode, corresponding to $Q_0=20\times10^6$. 
Finally, as the threshold power for soliton formation scales as $1/Q^2$, such high microresonator $Q$ allows soliton formation of 40 GHz repetition rate with only 10 mW optical power, without using an optical power amplifier (The measured GVD is $D_2/2\pi\sim224$ kHz). 

Next, we demonstrate wafer-scale yield of our fabrication technology.  
Figure \ref{Fig:Fig2}(b) shows our mask layout comprising $4\times4$ chip designs on the DUV stepper reticle.
Each chip has a $5\times5$ mm$^2$ size, and contains multiple microresonators as shown in Fig. \ref{Fig:Fig1}(b).
The DUV stepper writes the reticle pattern uniformly over the full 4-inch wafer in discrete fields. 
The calibration chips of 40 GHz FSR studied here are the C7 chips. 
The most probable value of $\kappa_0/2\pi$ histograms of C7 chips is measured and plotted in each exposure field, as shown in Fig. \ref{Fig:Fig2}(b). 
In most fields, $\kappa_0/2\pi\leqslant 7.5$ MHz is found.
While exceptionally narrow linewidth has been reported previously on individual resonances, our statistics based on tens of thousands of analyzed resonances from dozens of samples at different wafer positions, shows wafer-scale fabrication throughput and yield.

In addition, sideband modulation technique \cite{Li:12} is performed to measure the resonance linewidth $\kappa/2\pi$ and to fit $\kappa_0/2\pi$. 
Two sidebands, each separated from the carrier by 100 MHz, are used to calibrate the resonance linewidth. 
Figure \ref{Fig:Fig2}(c) compares the measured $\kappa/2\pi$ and fitted $\kappa_0/2\pi$ of the same resonance which does not present a visible mode split, using both the frequency-comb-assisted diode laser spectroscopy ($\kappa/2\pi=7.75$ MHz and $\kappa_0/2\pi=5.87$ MHz) and the sideband modulation technique ($\kappa/2\pi=7.95$ MHz and $\kappa_0/2\pi=6.05$ MHz). 
Both methods agree with each other, and show $Q_0>32\times10^6$.

Furthermore, cavity ring-down measurement is performed to validate the measured linewidth (see Method).
Figure \ref{Fig:Fig2}(d) shows the schematic of the experimental setup and a representative ring-down measurement data. 
The fitted optical field decay time is 37.8 ns, corresponding to 18.9 ns photon storage time.
The calculated loaded linewidth is $\kappa/2\pi=8.4$ MHz, showing consistency between the three characterization methods used here.

\begin{figure*}[t!]
\centering
\includegraphics[clip,scale=1]{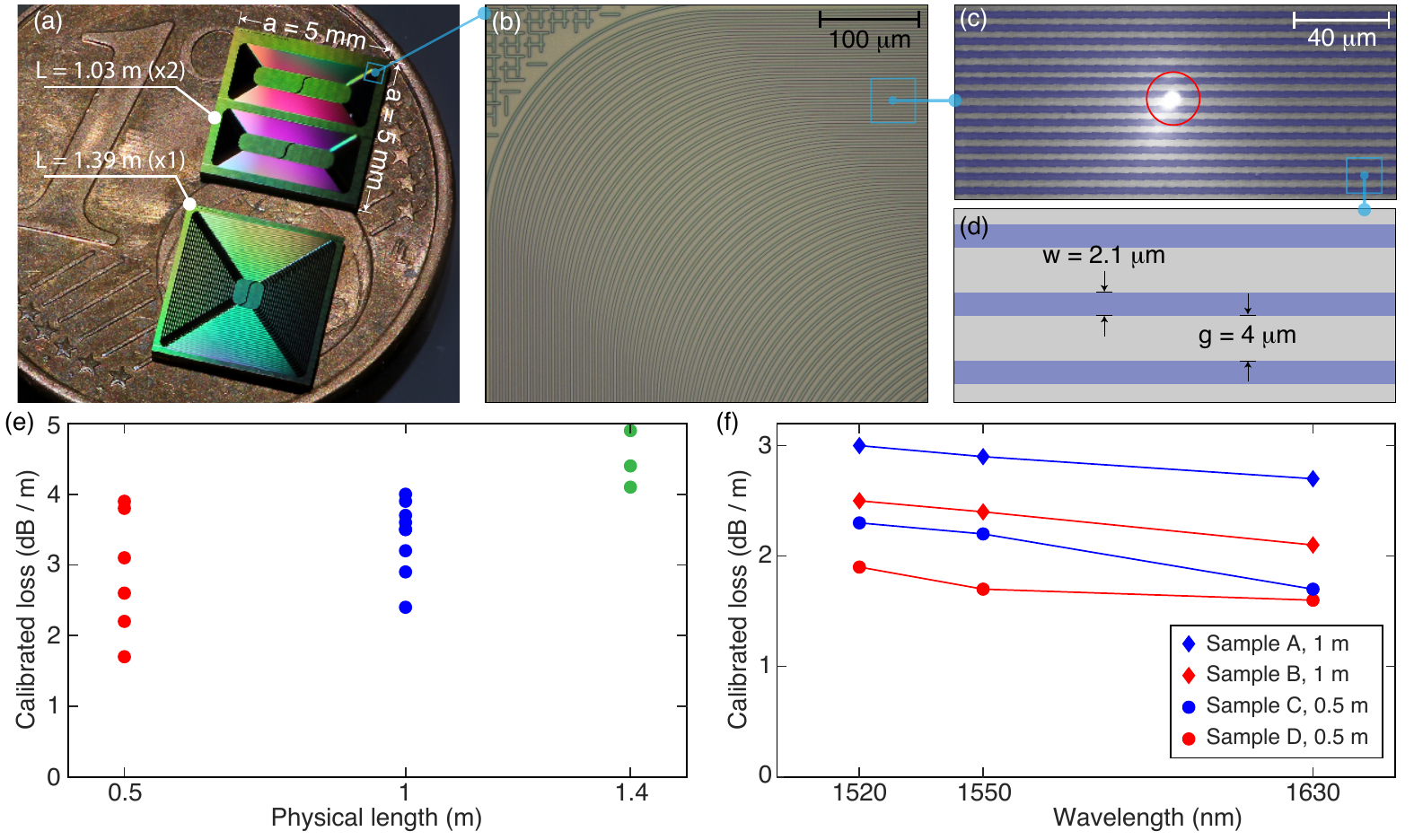}
\caption{
\textbf{Small-footprint, meter-long, ultralow-loss Si$_3$N$_4$ spiral waveguides}. 
(a)~Photograph showing Si$_3$N$_4$ chips containing two 1.0-meter-long and one 1.4-meter-long spiral waveguides.
(b, c)~Optical micrographs of the densely packed Si$_3$N$_4$ waveguides in Archimedean spirals, with yellow light camera (b) and IR camera (c).
When 1550 nm laser light is coupled into the waveguide, light-scattering defects are observed under the IR camera (highlighted with a red circle).
(d)~Schematic showing the waveguide width and spacing.
(e)~Measured and calibrated optical losses in 0.5 m, 1.0 m and 1.4 m long spiral waveguides. 
(f)~Measured  and calibrated losses measured at different wavelengths for four selected samples. 
}
\label{Fig:Fig3}
\end{figure*}

\noindent \textbf{Meter-long spiral waveguides}:
In addition to high-$Q$ microresonators, we also fabricate and characterize meter-long spiral waveguides that are key elements to build photonic true-time delay lines. 
Previously, silica suspended wedge waveguides \cite{Lee:12a} and thin-core Si$_3$N$_4$ waveguides \cite{Bauters:11} have been studied to build delay lines with losses below 0.1 dB/m. 
However, as a result of avoiding bending losses, these waveguides occupy more than 20 cm$^2$ areas, thus suffering from significant device footprints. 
While tight optical confinement can reduce the footprint, losses approaching even 1 dB/m have not been achieved in any nonlinear waveguide including thick-core Si$_3$N$_4$.
Here, we demonstrate meter-long Si$_3$N$_4$ waveguides featuring ultralow loss and small footprint, which can enable key applications for travelling-wave parametric amplifiers \cite{Foster:06, Kuyken:11, Ooi:17, YangM:18}, rare-earth-doped mode-locked lasers \cite{Xin:19} and optical coherence tomography (OCT) \cite{Ji:19a}.

Figure \ref{Fig:Fig3}(a) shows a photograph of photonic chips containing Si$_3$N$_4$ waveguides of physical lengths $L$ longer than 1 m. 
Figure \ref{Fig:Fig3}(b, c, d) shows the spiral layout.
The waveguides are densely packed in Archimedean spirals, with waveguide width $w=2.1$ $\mu$m and gap distance $g=4$ $\mu$m.
Three lengths are studied here: 
a 0.5-meter-long spiral contains 50 coils and covers 3.1 mm$^2$ area; 
a 1.0-meter-long spiral contains 106 coils and covers 6.6 mm$^2$ area; 
and a 1.4-meter-long spiral contains 130 coils and covers 20.2 mm$^2$ area. 
Compared with the previous report based on thin-core Si$_3$N$_4$ waveguides \cite{Bauters:11} showing a device footprint of more than 20 cm$^2$ area for 1 m physical length, our devices represent a footprint reduction of 300 times, critical for photonic integration. 
Figure \ref{Fig:Fig3}(e) shows the measured losses in multiples samples, calibrated using the adjacent 5-millimeter-long waveguide which has a fiber-chip-fiber through coupling efficiency of 33\% (4.8 dB for two chip facets).
The lowest loss values found are 1.7 dB/m for 0.5 m length, 2.4 dB/m for 1.0 m length, and 4.1 dB/m for 1.4 m length.
These loss values are higher than the value extrapolated from microresonator $Q$ characterization (1.0 dB/m). 
Meanwhile, the overall trend shows higher losses in longer waveguides.
We attribute both observations to the extra light-scattering defects. 
Light-scattering defects are found under an infrared (IR) microscope, as shown in Fig. \ref{Fig:Fig3}(c).
By counting the number of defects in high-loss spirals, we estimate that each defect causes 1--2 dB extra loss. 
The probability of defects depends on the waveguide area.
These defects are likely caused by particle contamination on the wafer, as we have verified that these defects are not on the DUV reticle which would generate the same defects in the same position in each exposure field.
Figure \ref{Fig:Fig3}(f) shows the calibrated losses measured at different wavelengths for four selected samples. 
A trend showing a higher loss at a shorter wavelength is observed.

\begin{figure*}[t!]
\centering
\includegraphics[clip,scale=1]{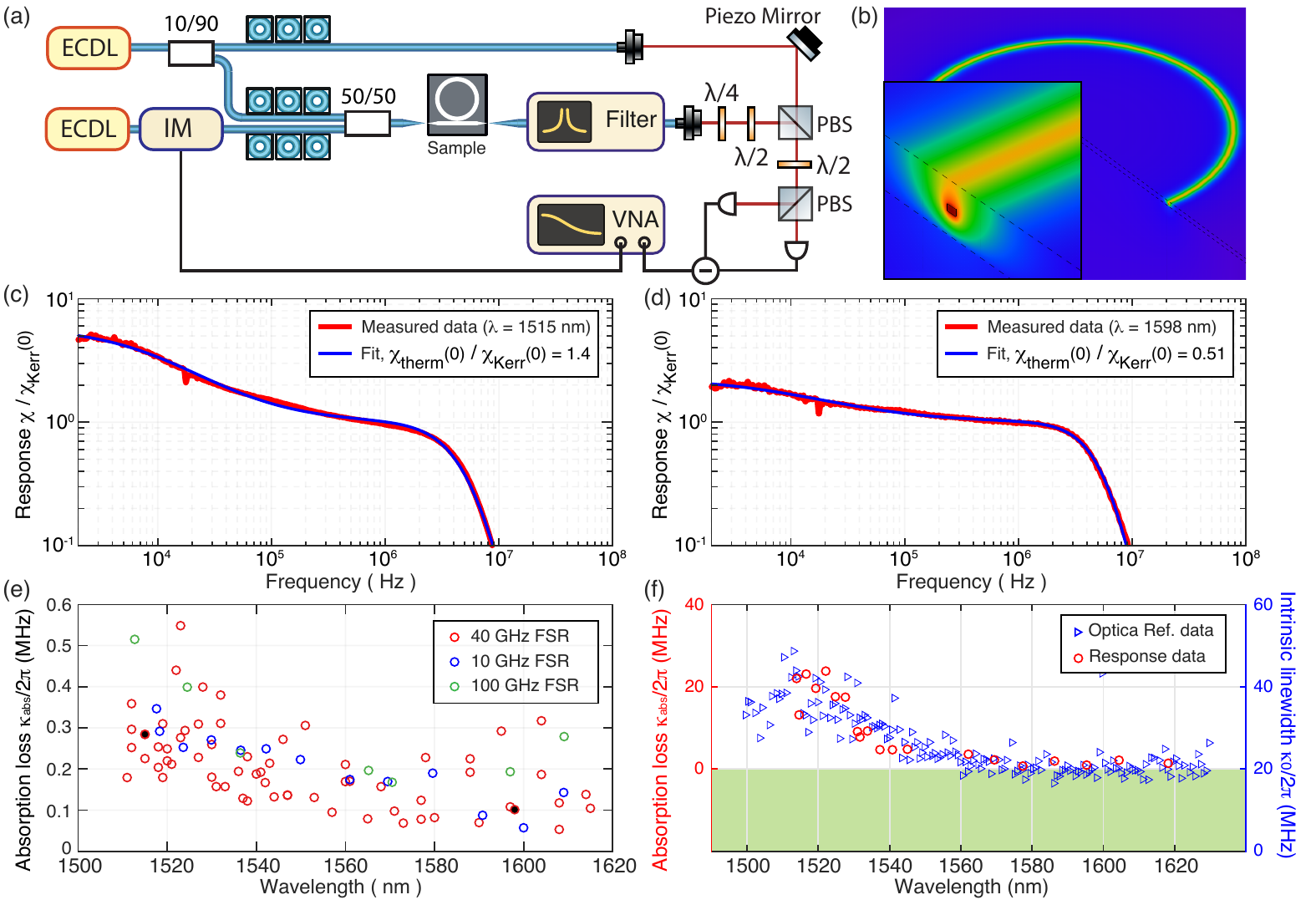}
\caption{
\textbf{Probing the ultimate absorption loss limit of Si$_3$N$_4$ microresonators via Kerr-nonlinearity-calibrated thermal response measurements}. 
(a) Experiment setup. 
ECDL: external-cavity diode lasers. 
IM: intensity modulator.
VNA: vector network analyzer.
PBS: polarization beam splitter.
(b) Thermal simulation of the temperature distribution in the waveguide structures. 
(c, d) Measured frequency response $\chi(\omega)$ normalized to $\chi_\mathrm{Kerr}$, of two representative resonances at 1515 nm and 1598 nm. 
The fitted thermal cutoff frequency $\omega_\mathrm{therm}/2\pi$ and cavity cutoff frequency $\kappa/4\pi$ are, $\omega_\mathrm{therm}/2\pi=14.2$ kHz and $\kappa/4\pi=6.2$ MHz in (c), and $\omega_\mathrm{therm}/2\pi=14.9$ kHz and $\kappa/4\pi=6.5$ MHz in (d). 
(e) Calculated absorption loss $\kappa_\mathrm{abs}/2\pi$ of different resonances from different samples.
The black solid circles correspond to the data shown in (c, d).
(f) Comparison of loss values measured using the linear response measurement and the frequency-comb-assisted diode laser spectroscopy, on a partially annealed sample.
This particular samples features prominent hydrogen absorption losses.
The green-shaded zoom marks a wavelength-independent scattering loss of 20 MHz.
} 
\label{Fig:Fig6}
\end{figure*}

\noindent \textbf{Quantitative analysis of loss limit:}
Next, we investigate quantitatively the intrinsic absorption and scattering losses of our Si$_3$N$_4$ waveguides.
The optical losses in the telecommunication band have two main contributions: the Rayleigh scattering loss caused mainly by the waveguide sidewall roughness, and the absorption loss due to e.g. hydrogen impurities. 
While the hydrogen absorption loss can be efficiently eliminated via repeated thermal annealing of Si$_3$N$_4$ at high temperature $\sim1200^\circ$C \cite{Luke:15, Liu:18a}, efforts on loss reduction have mainly focused on reducing waveguide roughness via optimized dry etching \cite{Ji:17}, wet etching \cite{HuangYW:15}, and etchless process \cite{Griffith:12, Han:19b}. 
In addition, the large mode area of thin-core Si$_3$N$_4$ waveguides \cite{Bauters:11, Gundavarapu:19, Spencer:14} results in reduced optical mode interaction with waveguide sidewall roughness, and thereby reduced scattering losses. 

To quantify the thermal absorption loss of our Si$_3$N$_4$ waveguides, a modulation response measurement \cite{Wilson:20} is performed. 
The experimental setup is shown in Fig. \ref{Fig:Fig6}(a), with two lasers, the pump and probe.
The pump laser is tuned to an optical resonance whose frequency is $f_m$, and the thermal absorption loss $\kappa_\mathrm{abs}$ in this resonance is to be characterized.
Meanwhile, the pump laser is intensity-modulated with frequency $\omega$.
The probe laser is loosely locked (i.e. low-bandwidth locking) to another optical resonance whose frequency is $f_m'$.
The principle of the linear microresonator response measurement is to characterize the resonance frequency shift $\delta f_{m'}=\chi(\omega)\delta n_\mathrm{ph}$ of the probe mode $f_{m'}$ induced by the intensity modulation of the pump mode $f_m$. 
This intensity modulation causes intracavity power modulation (i.e. photon number modulation $n_\mathrm{ph}$), which modulates the resonance frequency of the probe mode $f_{m'}$ via Kerr and thermal nonlinearities.
The pump power is maintained sufficiently low, such that the steady-state frequency shift of the probe mode is small compared to the resonance linewidth $\kappa$, i.e. $\delta f_{m'}\ll\kappa$. 
In this linear regime, the frequency response to the modulating pump power is given by \cite{Wilson:20}
\begin{equation}
\chi(\omega)=\frac{\delta f_{m'}}{\delta n_\mathrm{ph}}=\chi_\mathrm{therm}(\omega)+\chi_\mathrm{Kerr}(\omega)
\label{eq:chi}
\end{equation}

The total response $\chi(\omega)$ consists of two parts: the Kerr response $\chi_\mathrm{Kerr}(\omega)$ with infinite bandwidth, and the thermal response $\chi_\mathrm{therm}(\omega)$ with a bandwidth below 20 kHz.
Therefore, by calibrating the response $\chi(\omega)$ as a function of the modulation frequency $\omega$, $\chi_\mathrm{therm}(\omega)$ and $\chi_\mathrm{Kerr}(\omega)$ can be individually identified.
Using the values of $\chi_\mathrm{therm}(\omega)$ and $\chi_\mathrm{Kerr}(\omega)$ at DC ($\omega=0$), the absorption rate is calculated as 
\begin{equation}
\kappa_\mathrm{abs}=\frac{2cn_\text{mat}n_2}{n_g^2V_\mathrm{eff}\frac{dT}{dP_\mathrm{abs}}\frac{dn_\text{mat}}{dT}}\frac{\chi_\mathrm{therm}(0)}{\chi_\mathrm{Kerr}(0)}
\label{eq:kabs}
\end{equation}
where $V_\mathrm{eff}$ is the effective optical mode volume, $n_2=2.4\times10^{-19}$m$^2/$W is the nonlinear index of Si$_3$N$_4$, $n_g=2.1$ is the group index, $n_\text{mat}=2.0$ is the material index and $dn_\text{mat}/dT=2.5\times10^{-5}/$K is the thermo-optic coefficient \cite{Arbabi:13}, and $P_\mathrm{abs}$ is the absorbed power.

The frequency response $\delta f_{m'}$ to the pump modulation is transduced into the probe laser's phase modulation.
The phase response is measured using a balanced homodyne detection, with the pump laser being filtered out before detection (see Methods).
To evaluate the absorption rate $\kappa_\mathrm{abs}$, the factor $\chi_\mathrm{therm}(0)/\chi_\mathrm{Kerr}(0)$ is retrieved by a two-pole fitting of the measured response $\chi(\omega)$, which presents a thermal cutoff frequency $\omega_\mathrm{therm}$ and a cavity cutoff frequency $\kappa/2$. 
The fitting exploits the fact that the thermal response $\chi_\mathrm{therm}(\omega)$ dominates at frequency below 10 kHz and has a cutoff frequency $\omega_\mathrm{therm}/2\pi<20$ kHz. 
At higher frequency, the Kerr response $\chi_\mathrm{Kerr}(\omega)$ dominates.
Figure ~\ref{Fig:Fig6}(c, d) present two examples of measured and fitted $\chi(\omega)$.
Finite-element simulations of optical mode profiles and bulk absorption heating are performed to calculate the coefficients $V_\mathrm{eff}$ and $dT/dP_\mathrm{abs}$.
Figure \ref{Fig:Fig6}(b) shows the temperature profile from the thermal simulation (see Method).

Figure~\ref{Fig:Fig6}(e) shows the calculated absorption rates $\kappa_\mathrm{abs}$ of different resonances from four 40-GHz-FSR Si$_3$N$_4$ samples featuring $Q_0>30\times10^6$, in comparison with 10- and 100-GHz-FSR samples fabricated using the same process but from different wafers. 
All samples show similar trends, and present two conclusions.
First, the mean absorption loss is only $\kappa_\mathrm{abs}/2\pi\approx0.2$ MHz, corresponding an absorption-loss-limited $Q$ factor of $10^9$.
Therefore, the optical losses of our Si$_3$N$_4$ waveguides ($\kappa/2\pi=6.5$ MHz) are currently dominated by scattering losses. 
Second, $\kappa_\mathrm{abs}/2\pi$ is higher ($\approx0.4$ MHz) around 1520 nm, compared to the value at e.g. 1600 nm ($<0.2$ MHz). 
This is caused by the residual hydrogen impurities in our thermally annealed Si$_3$N$_4$. 
Note that, only standard LPCVD Si$_3$N$_4$ / SiO$_2$ films and thermal annealing are used in our fabrication to achieve such low absorption losses.

To validate our findings, we further benchmark the linear response measurement by characterizing a partially annealed Si$_3$N$_4$ sample whose resonance linewidth data have been published in ref.~\cite{Liu:18a}.
We characterize again this particular sample, using both the response measurement and the frequency-comb-assisted diode laser spectroscopy, and compare the results using both methods in Fig. \ref{Fig:Fig6}(f).
Assuming a wavelength-independent scattering loss of 20 MHz, the measured hydrogen absorption loss using the response measurement agrees with the total loss measured using the other method.

\noindent \textbf{Conclusion}: 
We have demonstrated a fabrication technology enabling high-yield and reproducible wafer-scale manufacturing of ultralow-loss, high-confinement, anomalous-GVD Si$_3$N$_4$ PIC.
We present a statistical study of microresonator losses based on tens of thousands of analyzed resonances.
We further reveal that our waveguide losses are dominated by scattering losses, which could be further reduced via e.g. optimized lithography and etching.
In the ideal case limited only by the thermal absorption loss, the potential microresonator $Q$ is calculated to exceed $10^9$ (corresponding to a linear loss of 0.03 dB/m).
The optimized photonic Damascene fabrication technology allows tight-confinement, ultralow-loss, high-yield, meter-scale, nonlinear PIC, and is suitable for adoption in CMOS foundries.

\noindent \textbf{Methods}

\medskip
\begin{footnotesize}
\noindent \textbf{Cavity ringdown}:
An intensity modulator (IM) is used to rapidly switch on and off the pump field. 
The ring-down signal of the transmitted light is recorded by a 1-GHz-bandwidth low-noise photodetector. 
A 50-kHz square wave electrical drive signal is generated using a fast arbitrary waveform generator, ensuring that the light is switched off significantly faster than the resonance linewidth. 
The upper and lower voltage levels of the square wave are adjusted to match the maximum and minimum transmission voltage of the IM, such that the electrical overshoot and undershoot of the square wave signal do not twist the ring-down slope. 
Due to the finite extinction ratio of the IM, the residual pump field beats with the leakage of the intracavity field, producing a field ring-down signal which is affected by the detuning of the laser from the cavity mode resonances~\cite{wojtewicz2018response}. 
At small detunings ($\kappa \gg\Delta$), the effective ring-down rate is increased by the laser's detuning from cavity resonance, and thus the directly inferred quality factor is less accurate than the sideband fitting result. 
Therefore, the ring-down results can only serve as a lower bound of the loaded $Q$ factor of the measured resonances. 
The estimated loaded linewidth $\kappa/2\pi=8.4$ MHz is in agreement with the sideband fitting results, showing consistency between the three characterization methods used here.

\noindent \textbf{Thermal simulations}:
We use COMSOL Multiphysics to simulate the thermal response due to bulk absorption heating of our Si$_3$N$_4$ samples.
The main material property coefficients of interest used in the current simulation are identical to the ones used in ref.\cite{Huang:19} for simulating the Si$_3$N$_4$ thermal refractive noise. 
We first simulate the waveguide optical mode profile (TE$_{00}$ mode), from which the effective mode volume $V_\mathrm{eff}$ is calculated.
Bulk absorption heating is introduced whose power distribution is proportional to the intensity distribution of the optical mode $f_m$.
From the stationary study of the sample heating, the dependence of temperature change on absorbed power, $dT/dP_\mathrm{abs}$, is retrieved from an absorption power sweep.
The combined value of $V_\mathrm{eff}\cdot dT/dP_\mathrm{abs}$ is calculated as $2.09\times \SI{e-12}{K\cdot m^3/W}$ in the case of full SiO$_2$ cladding for samples used in Fig. \ref{Fig:Fig6}(c, d, e), and is $3.84\times \SI{e-12}{K\cdot m^3/W}$ in the case of missing top SiO$_2$ cladding for samples used in Fig. \ref{Fig:Fig6}(f).

\noindent \textbf{Response calibration}:
In order to extract the actual microresonator response $\chi(\omega)$ from the experimentally photodetected $\chi'(\omega)$, the frequency response $\chi_\mathrm{det}(\omega)$ of our entire experiment setup and detection chain needs to be calibrated first.
This is realized by direct detection of the pump power modulation $\delta P(\omega)\propto\chi_\mathrm{det}(\omega)$ in the absence of the probe laser and the pump filter. 
The measured response $\chi'(\omega)$ is normalized to the setup response $\chi_\mathrm{det}(\omega)$, and thus the actual microresonator response $\chi(\omega)=\chi'(\omega)/\chi_\mathrm{det}(\omega)$ is retrieved, with an uncertain constant factor.
This constant factor is removed when retrieving $\chi_\mathrm{therm}(0)/\chi_\mathrm{Kerr}(0)$ from the two pole fitting of $\chi(\omega)$ using a fitting function
\begin{equation*}
\chi(\omega)=\frac{\chi_\mathrm{Kerr}(0)}{\sqrt{1+(2\omega/\kappa)^2}}\cdot (1+\frac{\chi_\mathrm{therm}(0)}{\chi_\mathrm{Kerr}(0)}\frac{1}{1+i(\omega/\omega_\mathrm{therm})^\gamma})
\end{equation*}
with $\omega_\mathrm{therm}/2\pi$ and $\kappa/4\pi$ being the thermal and cavity cutoff frequencies, $\gamma$ being the parameter accounting for the material inhomogeneity (that is, the Si$_3$N$_4$ waveguide has a finite dimension and is surrounded by SiO$_2$ cladding).
In Fig.~\ref{Fig:Fig6}(c, d), only the normalized response $\chi(\omega)/\chi_\mathrm{Kerr}(0)$ is shown, with the uncertain constant factor removed.

\noindent \textbf{Funding Information}: This work was supported by Contract HR0011-15-C-055 (DODOS) from the Defense Advanced Research Projects Agency (DARPA), Microsystems Technology Office (MTO), by the Air Force Office of Scientific Research, Air Force Materiel Command, USAF under Award No. FA9550-15-1-0250, and by Swiss National Science Foundation under grant agreement No. 176563. (BRIDGE).


\noindent \textbf{Acknowledgments}: We thank Bahareh Ghadiani for the assistance in the fabrication process development in the early stage, and Qi-Fan Yang for the fruitful discussion. The Si$_3$N$_4$ microresonator samples were fabricated in the EPFL center of MicroNanoTechnology (CMi).

\noindent \textbf{Data Availability Statement}: The code and data used to produce the plots within this work will be released on the repository \texttt{Zenodo} upon publication of this preprint.

\end{footnotesize}
\bibliographystyle{apsrev4-1}
\bibliography{bibliography}
\end{document}